\documentclass[12pt]{article}
\usepackage{amsfonts}
\usepackage{latexsym,amsmath}
\usepackage{amssymb,array,graphicx}
\usepackage{float}
\usepackage{booktabs}
\usepackage{enumerate}
\usepackage{rotating}
\usepackage{geometry}
\usepackage{color}
\begin{document}
\title{A new approach of chain sampling inspection plan}
\author{Harsh Tripathi$^{a}$ and Mahendra Saha $^{a}$\footnote{Corresponding author e-mail: mahendrasaha@curaj.ac.in}\\ 
\small $^{a}$ Department of Statistics, Central University of Rajasthan, Rajasthan, India}
\date{}
\maketitle
\begin{abstract}
To develop decision rules regarding acceptance or rejection of production lots based on sample data is the purpose of acceptance sampling inspection plan. Dependent sampling procedures cumulate results from several preceding production lots when testing is expensive or destructive. This chaining of past lots reduce the sizes of the required samples, essential for acceptance or rejection of production lots. In this article, a new approach for chaining the past lot(s) results proposed, named as modified chain group acceptance sampling inspection plan is more effective than widely used plan. Several properties of operating characteristic curves are derived. A comparison study has been done between the proposed and group acceptance sampling inspection plan as well as single acceptance sampling inspection plan. A example has been given to illustrate the proposed plan in a good manner.
\end{abstract}
\section*{Abbreviations:}
\begin{tabular}{lcl}
ASIP &:& Acceptance sampling inspection plan.\\
SASIP &:& single acceptance sampling inspection plan.\\
DASIP &:& Double acceptance sampling inspection plan.\\
GASIP &:& Group acceptance sampling inspection plan.\\
SEASIP &:& Sequential acceptance sampling inspection plan.\\
ChSP &:& Chain sampling plan.\\
MChSP &:& Modified chain sampling plan.\\
MChGSP &:& Modified chain group sampling plan.\\
OC &:& Operating characteristic.\\
\end{tabular}

\section{Introduction}
ASIP is classifies in two broad areas: sampling inspection plan by attribute and sampling inspection plan by variable. In literature many attribute sampling inspection plans are available viz., SASIP, DASIP, GASIP, SESAIP etc., while variable sampling plan uses the accurate measurements of quality characteristics for decision-making rather than classifying the products as conforming or non-conforming. Both type (attribute and variable) sampling inspection plan are used for sentencing a lot based on sample of that lot. Plan parameters of both (attribute and variable) sampling inspection plan are determined with the help of two point approach.\\

Many researcher have discussed the time truncated SASIP and some of them listed here, namely, Gupta$(1962)$, Gupta et al.$(1961)$, Rosaiah et al. $(2005)$, Tsai et al. $(2006)$, Baklizi et al. $(2004)$, Balakrishnan et al.$(2007)$, Aslam et al. $(2010)$ and Al-omari $(2015)$. Many authors have discussed the time truncated DASIP namely, GS Rao $(2011)$, Ramaswamy et al. $(2012)$ and Gui $(2014)$. Also, many researchers have studied GASIP with time truncated life test and readers may refer to Aslam et al. ($2009$) for gamma distribution, Aslam et al. ($2009, 2011$) for Weibull and Birnbaum-Saunders distributions, Rao ($2011$) for Marshall-Olkin extended exponential distribution.\\

ChSP first introduced by Dodge $(1955)$, also known as ChSP-1 plan. ChSP-1 is a plan with zero acceptance number sampling inspection plan and developed for the inspection by attribute as well as by variable [see, Govindaraju $(2006)$ and Govindaraju, Balamurali $(1998)$]. ChSP-1 inspection plan depends on chain past lot results, i.e., quality of past lot or inspection of past lot plays an important role in the decision making process of sentencing a lot. Balamurali and Usha $(2013)$, Govindaraju and Subramani $(1993)$ have done the computation of tables and results considering ChSP-1 plan. In ChSP-1 uses past results only when a non-conforming unit is observed in current sample. Govindraju and Lai $(1998)$ have developed a modified version of ChSP-1 plan, is known as MChSP-1 plan. However the MChSP-1 plan can only be used for inspection by attributes and their selection is only studied under the condition of a Poisson model. Now Luca $(2018)$ has developed an extension of MChSP-1 plan, known as modified sampling (MChSP) plan. MChSP is applicable in both attribute and variable inspection. \\
In this article we have developed a new sampling inspection plan which is combination of MChSP and ordinary GASIP sampling inspection plan, named as MChGSP. Proposed plan is applicable in case of ASIP by attribute. Rest of the article is organized as follows: In section $2$, design of the propose plan is discussed. Description of tables and the comparison of the proposed plan with GASIP and SASIP  are discussed in section $3$. In section $4$, we have given a example to understand the methodology and the applicability of the proposed plan. Finally, concluding words about the findings of the proposed study is placed in section $5$. \\

\section{Design of modified chain group sampling plan} 
In this section, we have proposed a new sampling inspection plan, named as MChGSP. Plan parameters of MChGSP are the number of groups ($f$), acceptance number ($c$) and number of chained sample results ($i$) respectively. A MChGSP plan is determined by the triple of natural numbers ($f$, $c$, $i$).\\
Now, the procedure of MChGSP is follows:

\begin{enumerate}
\item Select $n$ items from a particular lot and allocate $e$ items to $f$ groups, i.e, $n=e \times f$. Start with normal inspection for pre-fixed experiment time $t$.
\item Inspect all the groups simultaneously and record the number of non-conforming units (d) upto pre-fixed experiment time.
\item Go to MChSP inspection plan, If $d \le c$ the lot is accepted provided that there is at-most 1 lot among the preceding i lots in which the number of defective units d exceeds the criterion c, otherwise reject the lot.
\end{enumerate}

Now, the probability of acceptance of GASIP is obtained by the following Equation:

\begin{eqnarray}\label{eq1}
E=\sum\limits_{i=0}^{c}{ef \choose i} p^i (1-p)^{(ef-i)}
\end{eqnarray}
where, $p$ is the probability that observed number of failures occurs before the experimental time $t$ 

%
%
Expression of OC function of the proposed plan is given below: \\

$$
E_a(p)=\sum\limits_{i=0}^{c}{ef \choose i} p^i (1-p)^{(ef-i)}\left[\left(\sum\limits_{i=0}^{c}{ef \choose i} p^i (1-p)^{(ef-i)}\right)^i+i\left(\sum\limits_{i=0}^{c}{ef \choose i} p^i (1-p)^{(ef-i)}\right)^{(i-1)}\left(1-\sum\limits_{i=0}^{c}{ef \choose i} p^i (1-p)^{(ef-i)}\right))\right]
$$

Therefore, we can use two-point approach (at AQL and LQL) to determine the plan parameters of the proposed plan by using the following non-linear optimization problem:\\\\

\begin{eqnarray}\label{eq8}
\mbox{Minimize, ASN:}~~ n=f\times e
\end{eqnarray}

\begin{eqnarray}\label{eq9}
E_a(p_0)=E_0(E_0^i+iE_0^{i-1}(1-E_0)) \ge (1-\alpha)
\end{eqnarray}

\begin{eqnarray}\label{eq10}
E_a(p_1)=E_1(E_1^i+iE_1^{i-1}(1-E_1)) \le \beta
\end{eqnarray}

In above optimizing problem, our aim is to minimize sample size $n$ and $n$ depends on number of groups $g$ for given group size $r$, i.e., we have to minimize the  number of groups $g$ in such a way that $g$ satisfies above optimization problem for given $r$. 

\section{Description of Tables}
In this section, we have described the tables. Table $1$ represents the plan parameters of the proposed plan for the group size $(r=5)$ when consumer's risk $(\alpha=0.05)$ and producer's risk $(\beta=0.10)$ are prefixed and for the given value of AQL ($p_0$) and LQL ($p_1$). Number of groups decreases for the fixed AQL and varying values of LQL. Table $2$ shows the comparison study among the proposed plan, ordinary GASIP and SASIP. This comparison shows that the proposed plan perform better than ordinary GASIP and SASIP. In some cases, proposed plan required same number of groups as in ordinary GASIP and SASIP for the same set of values of AQL and LQL for sentencing the lot. We would prefer to use proposed plan over the ordinary GASIP and SASIP for the reason that past information plays significant role to take a decision about lot in proposed MChGSP plan rather than to take decision on the basis of current sample.

\section{Example}
Suppose that the producer's risk ($\alpha$) and consumer's risk ($\beta$) are $0.05$ and $0.10$ respectively. Values of AQL ($p_0$) and LQL ($p_1$) are $0.05$ and $0.14$ respectively which are known to experimenter to apply the two point approach for the estimation of the plan parameters of proposed plan. From table $1$, plan parameters are $f=13$, $c=6$ and $i=2$ for the prefixed group size $e=5$. Based on theses obtained plan parameters, MChGSP is:

\begin{itemize}
\item Select a sample of size $65$ from a submitted lot. Allocate $5$ items to $13$ groups, i.e, $n=r \times f$ and start with normal inspection.
\item Inspect all the groups simultaneously and record the number of non-conforming units (d).
\item Go to MChSP inspection plan, If $d \le 6$ the lot is accepted provided that there is at-most 1 lot among the preceding 2 lots in which the number of defective units d exceeds the $6$, otherwise reject the lot.
\end{itemize}

\section{Conclusions}
In this article, we have introduced a new sampling inspection plan, name as MChGSP. We compared the proposed plan with exist ordinary GASIP and SASIP. MChGSP provides flexibility to reach a decision regarding lot acceptance or rejection with the minimum number of groups by using past results.

\section{Reference}
\begin{enumerate}

\item Aslam, M., Kundu, D., Ahmed, M.(2010). Time truncated acceptance sampling plans for generalized exponential distribution,{\it Journal of Applied Statistics}, {\bf 37(4)}, 555-566.

\item A.I, Al-Omari.(2015). Time truncated acceptance sampling plans for generalized inverted exponential distribution,{\it Electronic Journal of Applied Statistical Analysis,}, {\bf 8(1)}, 1-12.

\item Aslam, M., Jun, C.H. and Ahmad, M. (2009). A Group sampling plan based on truncated life test for gamma distributed items. Pakistan Journal of Statistics. 2009; 25(3): 333-340.

\item Aslam, M., and Jun, C.-H. (2009). A group acceptance sampling plan for truncated life test having Weibull distribution. {\it Journal of Applied Statistics}, 36(9), 1021-1027.

\item Aslam, M., Jun, C.-H. and Ahmad, M. (2011). New acceptance sampling plans based on life test for Birnbaum-Saunders distributions. {\it Journal of Statistical Computation and Simulation}, 81(4), 461-470.
%

%
%
\item Balamurali, S. and Usha, M. (2013). Optimal designing of variables chain sampling plan by minimizing

\item Baklizi, A., EL Masri, A.E.K.(2004). Acceptance sampling plan based on truncated life tests in the Birnbaum Saunders model,{\it Risk Analysis}, {\bf 24}, 1453-1457.

\item Balakrishnan, N., Lieiva, V., Lopez, J.(2007). Acceptance sampling plan from truncated life tests based on generalized Birnbaum Saunders distribution,{\it Communication in Statistics-Simulation and Computation}, {\bf 34(3)}, 799-809.

\item Dodge, H.F. (1955). Chain Sampling inspection plan, {\it Industrial Quality Control}, {\bf 11(4)}, 10-13.
%

\item Govindaraju, R. (2006). Chain sampling, in Springer Handbook of Engineering Statistics, H. Pham, ed.,
Springer, London, 263–279.

\item Govindaraju, K. and Subramani, K. (1993). Selection of chain sampling plans ChSP-1 and ChSP-(0, 1)
for given acceptable quality level and limiting quality level, Amer. J.Math. Manag. Sci. 13 , 123–136.

\item Govindaraju, K. and Balamurali, S. (1998). Chain sampling plan for variables inspection, Journal of Applied Statistics. 25, 103–109.

\item Govindaraju, K. and Lai, C. D. (1998). A modified ChSP-1 chain sampling plan, MChSP-1, with very small sample size, Amer. J. Math. Manag. Sci. 18, 343-358.

\item Gupta, S.S(1962). Life test sampling plans for normal and lognormal distributions,{\it Technometrics}, {\bf 4(2)}, 151-175.

\item Gupta, S.S. and Groll, P.A.(1961). Gamma distribution in acceptance sampling based on life test,{\it Journal of the American Statistical Association}, {\bf 56(296)}, 942-970.

\item Gui, W.(2014). Double acceptance sampling plan for truncated life tests based on Maxwell distribution,{\it American Journal of Mathematical and Management Sciences}, {\bf 33}, 98-109.
%

\item Luca, S. (2018). Modified chain sampling plans for lot inspection by variable and attribute. Journal of Applied Statistics, 45(8), 1447-1464. 

\item Rosaiah, K. and Kantam, R.R.L.(2005). Acceptance sampling plan based on the inverse Rayleigh distribution,{\it Economic Quality Control,}{\bf 20(2)},77-286.

\item Rao, G.S.(2011). Double acceptance sampling plan based on truncated life tests for Marshall-Olkin Extended exponential distribution,{\it Austrian Journal of Statistics,}{\bf 40(3)}, 169-176.

\item Rao, G.S.(2011). Double acceptance sampling plan based on truncated life tests for Marshall-Olkin Extended exponential distribution,{\it Austrian Journal of Statistics}, 40(3), 169-176.

\item Sudamani Ramaswamy, A.R.(2012). Double acceptance sampling based on truncated life tests in generalized exponential distribution,{\it Applied Mathematical Sciences,}{\bf 6(64)}, 3199-3207.
%
%
%
%
%
%
\item Tsai, T.R., Wu, S.J.(2006). Acceptance sampling plan based on truncated life tests for generalized Rayleigh distribution,{\it Journal of Applied Satistics,}{\bf 33}, 595-600.
\end{enumerate}

\begin{table}[h]
\centering
\caption{The plan parameters (g,c,i) of MChGSP for r=5 and $\alpha=0.05$ and $\beta=0.10$.}
\begin{tabular}{lcccccccccccccccccccc}
 \hline
 AQL $(p_0)$ & LQL $(p_1)$ & $g$   &  c   & i   & $P_a(p_0)$  & $P_a(p_1)$   \\
\bottomrule
 $0.01$      & $0.02$       & $120$ & $10$ & $3$ & $0.9533257$ & $0.0949371$  \\
             & $0.03$       & $66$  & $7$  & $2$ & $0.9804377$ & $0.0900610$  \\
             & $0.04$       & $44$  & $6$  & $2$ & $0.9928176$ & $0.0861985$  \\
             & $0.05$       & $32$  & $4$  & $1$ & $0.9769802$ & $0.0938539$  \\
             & $0.06$       & $22$  & $3$  & $1$ & $0.9749619$ & $0.0980303$  \\
             & $0.07$       & $15$  & $2$  & $1$ & $0.9603309$ & $0.0967880$  \\
\bottomrule
\bottomrule
 $0.05$      & $0.10$       & $24$  & $10$ & $3$ & $0.9573947$ & $0.0883700$  \\
             & $0.12$       & $18$  & $8$  & $2$ & $0.962578$  & $0.0962938$  \\
             & $0.14$       & $13$  & $6$  & $2$ & $0.9549205$ & $0.0574218$  \\
             & $0.18$       & $10$  & $5$  & $1$ & $0.9622238$ & $0.0928591$  \\
             & $0.20$       & $8$   & $4$  & $1$ & $0.9519717$ & $0.0759145$  \\
\bottomrule
\bottomrule
 $0.10$      & $0.20$       & $12$  & $10$ & $3$ & $0.9624775$ & $0.0795959$  \\
             & $0.25$       & $9$   & $8$  & $3$ & $0.9670151$ & $0.0543979$  \\
             & $0.30$       & $7$   & $7$  & $2$ & $0.9796181$ & $0.0328563$  \\
             & $0.35$       & $5$   & $5$  & $2$ & $0.9666001$ & $0.0826247$  \\
             & $0.38$       & $4$   & $4$  & $1$ & $0.9568255$ & $0.0726116$  \\
\bottomrule
\bottomrule
 $0.15$      & $0.30$       & $8$  & $10$  & $3$ & $0.9675257$ & $0.07011739$  \\
             & $0.40$       & $5$   & $8$  & $3$ & $0.991839$  & $0.0543979$  \\
             & $0.50$       & $3$   & $5$  & $2$ & $0.9829121$ & $0.04209421$  \\
             & $0.55$       & $2$   & $3$  & $1$ & $0.9500302$ & $0.09084266$  \\
\bottomrule
    \end{tabular}
  \label{tab43 }
\end{table}
%


\begin{table}[h]
\centering
\caption{Comparison of proposed plan with existing GASIP and SASIP}
\begin{tabular}{lcccccccccccccccccccc}
 \hline
             &             & \multicolumn{2}{c}{r=5}                     &   \multicolumn{1}{c}{r=1} \\
     \hline
 AQL $(p_0)$ & LQL $(p_1)$ & MCh-GSP                & GASIP              &      SASIP                \\
             &             & $n=g\times r$          &$n=g\times r$       & $n=g\times r$             \\
\bottomrule
 $0.01$      & $0.02$       & $600=120 \times 5$     & $--$               & ---    \\
             & $0.03$       & $330=66\times 5$       & $395=79\times 5$   & $399$  \\
             & $0.04$       & $220=44\times 5$       & $325=65\times 5$   & $262$  \\
             & $0.05$       & $160=32\times 5$       & $195=39\times 5$   & $160$  \\
             & $0.06$       & $110=22\times 5$       & $135=27\times 5$   & $110$  \\
             & $0.07$       & $75=15\times 5$        & $80=16\times 5$    & $75$   \\
\bottomrule
\bottomrule
 $0.05$      & $0.10$       & $120=24\times 5$       & $--$               & ---    \\
             & $0.12$       & $90=18\times 5$        & $--$               & ---    \\
             & $0.14$       & $65=13\times 5$        & $--$               & ---    \\
             & $0.18$       & $50=10\times 5$        & $50=10\times 5$    & $50$   \\
             & $0.20$       & $40=8\times 5$         & $40=8\times 5$     & $40$   \\
\bottomrule
\bottomrule
 $0.10$      & $0.20$       & $60=12\times 5$        & $--$               & ---    \\
             & $0.25$       & $45=9\times 5$         & $--$               & ---    \\
             & $0.30$       & $35=7\times 5$         & $35=7\times 5$     & $41$   \\
             & $0.35$       & $25=5\times 5$         & $25=5\times 5$     & $25$   \\
             & $0.38$       & $20=4\times 5$         & $20=4\times 5$     & $20$   \\
\bottomrule
\bottomrule
 $0.15$      & $0.30$       & $40=8\times 5$         & $--$               & ---   \\
             & $0.40$       & $25=5\times 5$         & $30=65\times 5$    & $30$  \\
             & $0.50$       & $15=3\times 5$         & $--$               & $17$  \\
             & $0.55$       & $10=2\times 5$         & $--$               & ---   \\
\bottomrule
    \end{tabular}
  \label{tab49 }
\end{table} 

 \end{document}